\documentclass[10pt,showpacs, twocolumn]{revtex4-1}

\usepackage{graphicx}
\usepackage{amsmath}
\usepackage{csquotes}

\begin{document}

\title{ Effects of weak disorder on two-dimensional bilayered dipolar Bose-Einstein condensates}

\author{Abdel\^{a}ali Boudjem\^{a}a$^{1,2}$ and Redaouia Keltoum$^{1}$}

\affiliation{$^1$ Laboratory of Mechanics and Energy, \\
$^2$ Department of Physics, Faculty of Exact Sciences and Informatics, 
Hassiba Benbouali University of Chlef, P.O. Box 78, 02000, Chlef, Algeria.}

\email {a.boudjemaa@univ-chlef.dz}

\begin{abstract}

We investigate properties of two-dimensional bilayered dipolar Bose condensed gases in a weak random
potential with Gaussian correlation at zero temperature. 
Here the dipoles are oriented perpendicularly to the layers and in parallel/antiparallel configurations. 
We calculate analytically and numerically the condensate depletion, the one-body density-matrix, and the superfluid fraction
in the framework of the Bogoliubov theory. 
Our analysis not only provides fascinating new results do not exist in the literature 
but also shows that the intriguing interplay between the disorder, the interlayer coupling and the polarization orientation may lead to
localize/delocalize the condensed particles results in the formation of  glassy/superfluid phases. 
For a pure short-range interaction and a vanishing correlation length and a small interlayer distance, 
we reproduce the seminal results of Huang and Meng. 
While for a vanishing interlayer distance, our results reduce to the those obtained for single layer systems.


\end{abstract}


\maketitle

\section{Introduction}

Ultracold gases with dipole-dipole interaction (DDI) have attracted immense interest due to the long-range character and the anisotropy
in contrast to the short-ranged isotropic contact interaction  \cite{Baranov, Pfau,Carr, Pupillo2012}. 
In this respect, ultracold dipolar Bose gases have provided a tool for investigating complex many-body quantum effects. 
Among them one can quote, bifurcations, order, and chaos which have been analyzed using variational and numerical techniques \cite{Kobl, Mit, Rau, Gut}. 
Indeed, there are many nature inspired algorithms for solving the above complex nonlinear phenomena such as: ant colony optimization algorithm \cite{Deng},
fault diagnosis method  \cite{Zhao}, collaborative optimization algorithm \cite{Deng2},  and 
novel fault damage degree identification method based on high-order differential mathematical morphology gradient spectrum entropy \cite{Zhao1}.

The Bose-Einstein condensation (BEC) in a weak random external potential is ubiquitous in a large variety of systems. 
Recently, the interplay between interactions (contact and dipolar) and external disorder potentials in many-body systems 
has attracted a good deal of interest both theoretically and experimentally \cite{Clm,Schut, Lye, Clm1, Bily, Roat, Chen, Wht, HM,
Gior, Mish,Lugan2, Falco,Yuk, Lopa, Zob, Mor, Bhong,LSP, Lugan, Lugan1, Gaul, Gaul1,Lell, Krum,Nik, Ghab, Boudj,Boudj1,Boudj2, Boudj3,Boudj4, Boudj5}. 
It has been found that for a weak disorder, the density profile of the condensate follows the modulations of a smoothed random potential
while for very strong disorder the condensate decays into fragments in the disorder landscape. 
In the case of BEC with DDI,  the superfluid density acquires a characteristic direction dependence, 
i.e. the number of particles per volume participating in a superfluid motion varies with the chosen direction \cite{Krum,Nik, Ghab, Boudj,Boudj1}.
The interplay between the disorder and the rotonization in a quasi-two-dimensional (2D) dipolar BEC has been discussed by one of us \cite{Boudj2, Boudj3}.
Most recently, we have shown that the three-body interactions and the Lee-Huang-Yang quantum corrections 
play a crucial role in reducing the impacts of the disorder potential in BEC \cite {Boudj4, Boudj5}.

Over the past decade, ultracold dipolar gases in layered structures have attracted considerable attention
\cite {Wang, Wang1, Piko, Misha,  Shi, Pot,  Vol, Dalm, Santos1, Santos2, Ros, Klm, Fedo, Boudj6, Boudj7, Boudj8}.
Unlike single layers, these bilayered configurations in quasi-2D geometry exhibit many interesting phenomena namely: 
the formation of conventional and unconventional superfluids of polar molecules \cite {Wang, Piko, Misha,  Shi, Pot, Vol, Dalm, Fedo, Boudj6, Boudj7}, 
soliton molecules \cite{Santos1} and the enhacement of the roton instability \cite{ Wang1, Ros} due to the interlayer effects. 
The Fermi-polaron problem has been also discussed in such a bilayer system \cite{Klm}.
However, the contemporary problem of disordered ultracold dipolar bosons in bilayer systems has never been analyzed in the literature. 
Due to the availability of creating this bilayered configuration  experimentally by means of a 1D subwavelength lattice, 
it is then instructive  to study disordered BEC with DDI in bilayer arrangements. 
Such systems enable us to unveil the intriguing role of disorder, the interlayer effects and the dipolar interactions.

\begin{figure}
\centering{
\includegraphics[scale=0.3]{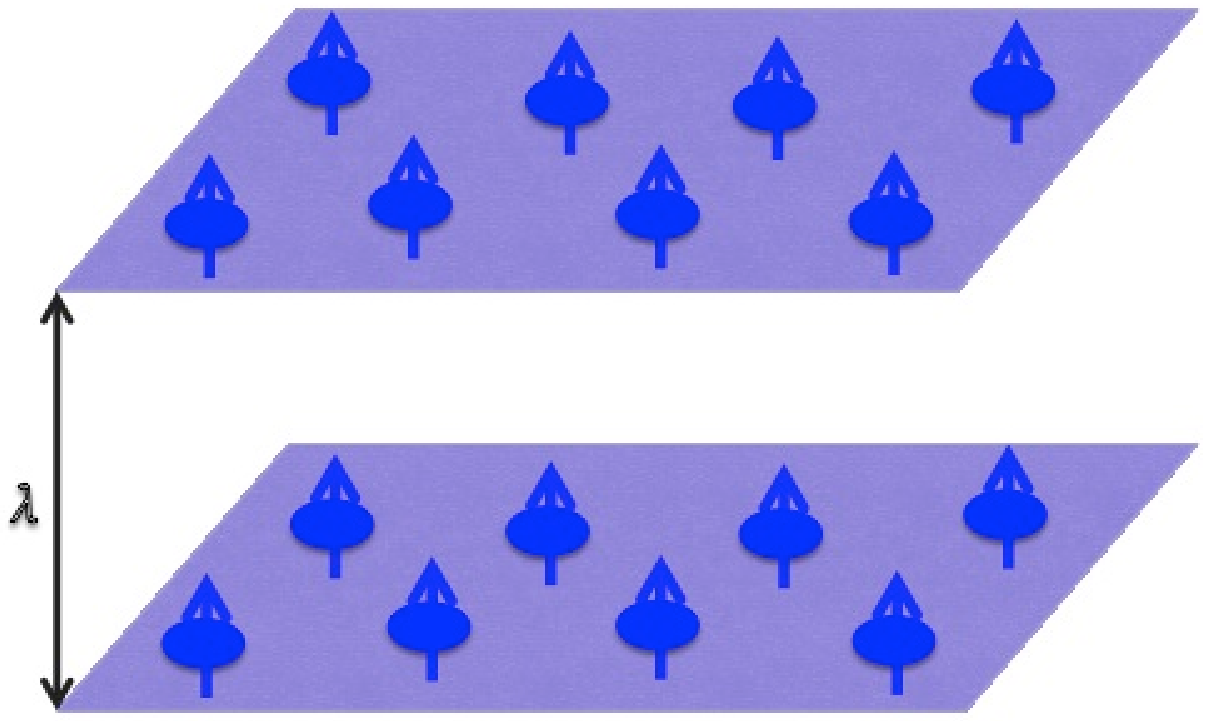}
\includegraphics[scale=0.3]{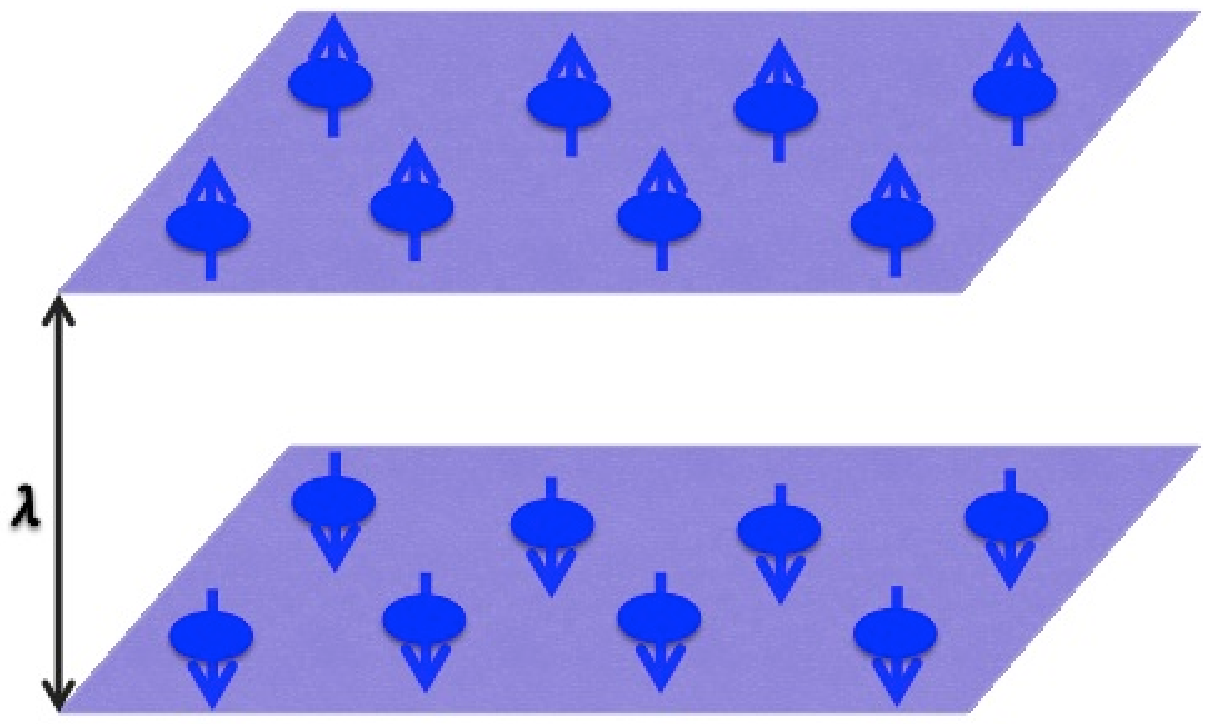}}
\caption{ (Color online) Schematic representaion  of the dipolar bilayer systems under consideration:
 (left) dipoles oriented in same directions in different layers (parallel configuration,  $\uparrow \uparrow$).
(right) dipoles oriented in opposite directions in different layers (antiparallel configuration, $\downarrow \uparrow$).}
\label{shemat}
\end{figure}

In this article, we investigate the problem of a disordered quasi-2D bilayered dipolar BEC with dipoles are oriented 
perpendicularly to the layers and in same (i.e parallel, denoted $\uparrow \uparrow$) /opposite (i.e antiparallel, denoted $\downarrow \uparrow$), directions 
in different layers (see Fig.\ref{shemat}).
To this end, we use the Bogoliubov-Huang-Meng  theory  \cite {HM}. 
Many studies have confirmed recently the effectiveness of this method in treating  dirty dipolar Bose gases \cite{Ghab, Boudj,Boudj1,Boudj2, Boudj3,Boudj4}.
We quantitatively examine the effects of varying polarization direction and interlayer DDI on
the collective excitations, glassy fraction, one-body density matrix and the superfluid fraction. 
Importantly, we find that in the parallel configuration, the interlayer DDI causes delocalization of particles enabling the transition to the superfluid phase.
Surprisingly, in the antiparallel arrangement, the bosons strongly fill the potential wells formed by disorder fluctuations 
depressing both the condensate and the superfluidity due to the intriguing interplay of the disorder and the interlayer DDI.
It is shown also that beyond a certain temperature depending on the polarization direction, the superfluid fraction vanishes.

The rest of the paper is organized as follows. 
In Sec.\ref{Mod} we introduce the Huang-Meng-Bogoliubov approach for a disordered bilayered dipolar  Bose gas.
The glassy fraction and the total quantum depletion are deeply analyzed. 
Section \ref{Coh} deals with the coherence of the system where we calculate numerically the one-body density matrix. 
We show that this quantity tends to a constant at large distance and it is almost insensitive to the interlayer distance.
In Sec.\ref{superf},  we shall analyze the role of DDI, interlayer coupling, polarization orientation and temperature on the superfluid fraction.  
Our conclusions are drawn in Sec.\ref{conc}.


\section{ Model} \label{Mod}

We consider a dilute Bose-condensed gas of dipolar bosons subjected to an external random potential loaded  in a quasi-2D bilayer setup.
Assuming vanishing hopping between layers and dipole moments $d$ are aligned perpendicularly to the plane of motion (cf. Fig\ref{shemat}). 
Our starting point is the secondly quantized Hamiltonian:
\begin{align}\label{ham}
\hat H\!\!=&\!\!\sum_j \bigg[\sum_{\bf k}\!E_k\hat a^\dagger_{j\bf k}\hat a_{j\bf k}\! +\!\frac{1}{S}\!\!\sum_{\bf k,\bf p} \! U_{j,\bf k\!-\!\bf p} \hat a^\dagger_{j\bf k} \hat a_{j\bf p} \\ \nonumber
&+\!\frac{1}{2S}\!\!\sum_{\bf k,\bf q,\bf p}\!\!
V_{jj}(\vert {\bf q}\!-\!{\bf p}\vert)\hat a^\dagger_{j,\bf k\!+\!\bf q} \hat a^\dagger_{j,\bf k\!-\!\bf q}\hat a_{j,\bf k\!+\!\bf p}\hat a_{j,\bf k\!-\!\bf p} , \\ \nonumber
&+\!\frac{1}{2S}\!\!\sum_{j'} \sum_{\bf k,\bf q,\bf p}\!\!
V_{jj'}(\vert {\bf q}\!-\!{\bf p}\vert)\hat a^\dagger_{j,\bf k\!+\!\bf q} \hat a^\dagger_{j',\bf k\!-\!\bf q}\hat a_{j',\bf k\!+\!\bf p}\hat a_{j,\bf k\!-\!\bf p} \bigg],
\end{align}
where $j=\pm 1$ is the layer index,  $S$ is the surface area, $E_k=\hbar^2k^2/2m$ is the energy of free particle, 
$\hat a_{\bf k}^\dagger$, $\hat a_{\bf k}$ are the creation and annihilation operators of particles, 
and $U$ is the disorder potential which is described by vanishing ensemble averages $\langle U(\mathbf r)\rangle=0$ and a finite correlation of the form 
$\langle U(\mathbf r) U(\mathbf r')\rangle=R (\mathbf r,\mathbf r')$.
In quasi-2D geometry, at large interparticle separations $r$ the intralayer interaction reads $V_{jj}(r)=d^2/r^3=\hbar^2r_*/mr^3$ \cite{Boudj9}, where 
$r_*=md^2/\hbar^2$ is the characteristic dipole-dipole distance, $d$ is the dipole moments,  and $m$ is the particle mass.
In momentum space it can be written as \cite{Boudj9}
\begin{equation}\label{ampl} 
 V_{jj}({\bf k})=g(1-C\vert {\bf k}\vert),
\end{equation}
where $g=g_{3D}/\sqrt{2}l_0$ is the 2D short-range coupling constant, $l_0=\sqrt{\hbar/m \omega}$, and $\omega$ is the confinement frequency, 
and $C =2\pi \hbar^2r_*/mg$. \\
The interlayer interaction potential $(j\neq j')$ is given by \cite {Piko, Misha, Ros, Klm, Fedo, Boudj6}
\begin{equation} \label{pot}
 V_{jj'}(r)=V_{\uparrow \uparrow , \downarrow \uparrow} (r) = \pm \, d^2\frac{r^2-2\lambda^2}{(r^2+\lambda^2)^{5/2}}.
\end{equation}
The potential $ V_{jj'}(r)$ is attractive at large/short distances $r$ depending on the dipoles orientation leading to the formation of an interlayer bound state.
The corresponding Fourier transform is given by
$V_{\uparrow \uparrow , \downarrow \uparrow} ({\bf k})  =\int d{\bf r} V_{\uparrow \uparrow , \downarrow \uparrow} ({\bf r})  e^{- i {\bf k r}}$.
After some algebra, we obtain for the two configurations:
\begin{equation}  \label{PFT}
V_{\uparrow \uparrow , \downarrow \uparrow} ({\bf k}) = \mp \frac{2\pi\hbar^2} {m} r_*|{\bf k}| e^{-|{\bf k}|\lambda},\\
\end{equation}
For  $k\lambda \ll 1$,  $V(k) = (2\pi\hbar^2/m) \,r_* k$.  
This linear dependence on $k$ originates from the so-called anomalous contribution to scattering \cite{Fedo,Boudj6}. 

\begin{figure}
\includegraphics[scale=0.45]{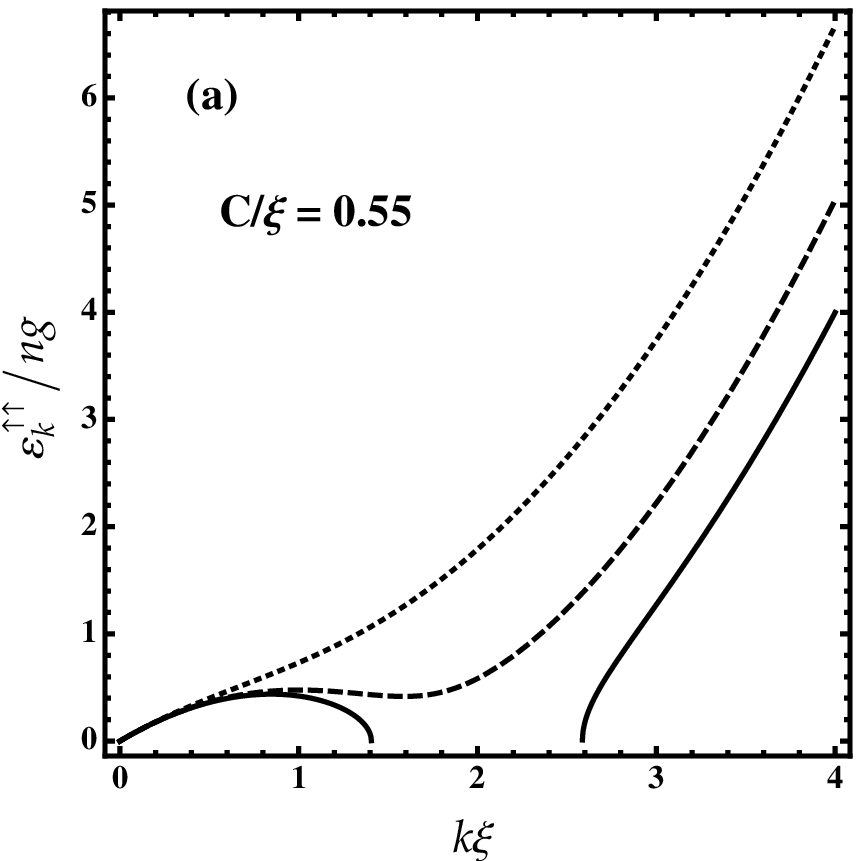}
\includegraphics[scale=0.46]{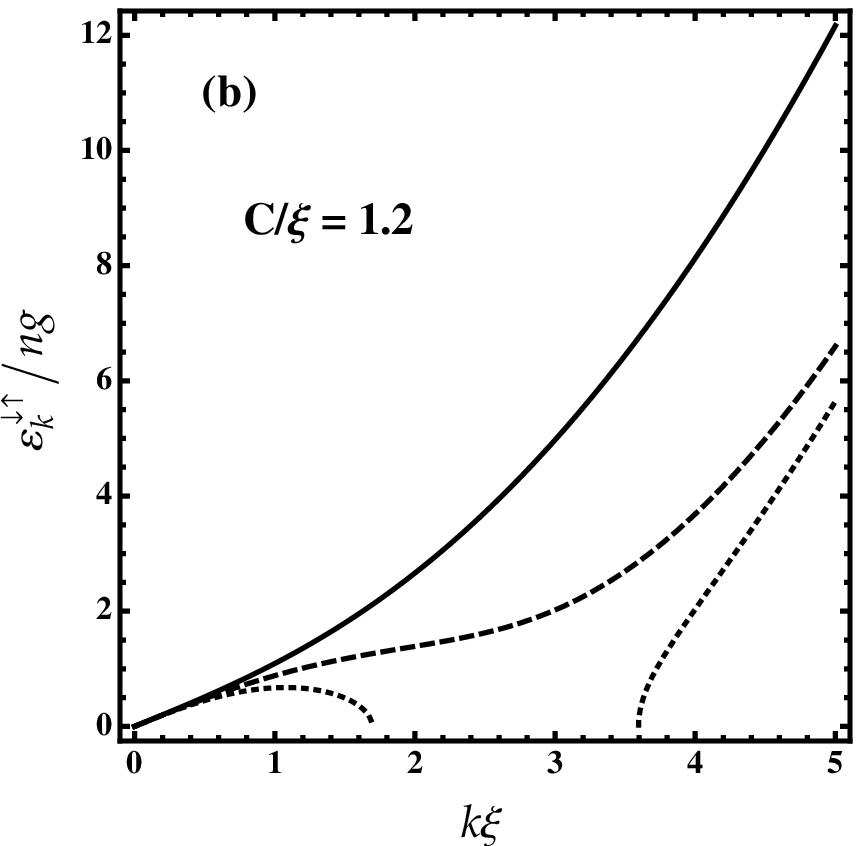}
\caption {The Bogoluibov excitations spectra  (a) $\varepsilon_{k\, \uparrow \uparrow}$  and (b) $\varepsilon_{k\, \downarrow \uparrow}$ 
 from Eq.(\ref{spectrum})  for several values of $\lambda$.
Solid line: $\lambda=0.05$. Dashed line: $\lambda=0.2$. Dotted line: $\lambda=1.1$.}
\label{disp}
\end{figure}

Now, we address the regime of weak interactions i.e. $mg/2\pi\hbar^2\ll 1$ and $r_*\ll \xi$, with $\xi=\hbar/\sqrt{mng}$ being the healing length,
and sufficiently weak external disorder potential.
The Hamiltonian (\ref{ham}) can be diagonalized using the Bogoliubov-Huang-Meng  transformation \cite {HM}: 
$\hat{a}_{k} =u_k \hat{b}_k-v_k\hat{b}_{-k}^\dagger -\beta_k$,
where $\hat b^\dagger_{\bf k}$ and $\hat b_{\bf k}$ are operators of elementary excitations.
The Bogoliubov functions $ u_k,v_k$ are expressed in a standard way: $ u_k,v_k=(\sqrt{\varepsilon_k/E_k}\pm\sqrt{E_k/\varepsilon_k})/2$, 
$\beta_{\bf k}=\sqrt{n/S} U_k E_k/\varepsilon_k^2$, where $S$ is the surface area.  The Bogoliubov excitations energy reads
\begin{equation}  \label{spectrum}
\varepsilon_{k\, \uparrow \uparrow , \downarrow \uparrow}=\sqrt{E_k^2+2ngE_k(1-Ck \mp C k  e^{-k\lambda})}, 
\end{equation}
For $k\lambda \gg 1$, the interlayer DDI vanishes and thus, the spectrum (\ref{spectrum}) 
reproduces analytically the roton-maxon structure seen in the 2D ordinary dipolar BEC (i.e. single layer) \cite{Boudj9}.  
For $k\lambda \ll 1$, one has $\varepsilon_{k\, \uparrow \uparrow }=\sqrt{E_k^2+2ngE_k(1- 2Ck)} $ which is similar to the single layer spectrum,  while 
$\varepsilon_{k\,\downarrow \uparrow}=\sqrt{E_k^2+2ngE_k}$ is equivalent to the spectrum of a nondipolar BEC.
One can conclude that for a bilayer system of dipoles with the antiparallel polarization of dipolar moments in two layers, the interlayer effects is 
important only for large enough interlayer distance $\lambda$ in stark contrast with the parallel configuration.
At low momenta $k\rightarrow 0$, the excitations are linear in $k$ (phonon regime) $\varepsilon_k= \hbar c_s k$, where 
$c_s=\sqrt{ng/m}$ is the sound velocity, it does not depend neither on the interlayer DDI nor on the intralayer DDI regardless the value of $\lambda$
and the polarization directions (see Fig.\ref{disp}). 
At higher momenta, it becomes quadratic as in the nondipolar case (see Fig.\ref{disp}) for any interlayer distance.
The dispersion relation changes it behavior and exhibits roton-maxon structure at intermediate $k$ as is shown in  Figs.\ref{disp}.(a) and (b).
The position and the energy of the roton strongly depend on the effect of varying polarization orientation and interlayer DDI.
For instance, the roton can be formed in the dispersion spectrum for very small $\lambda$ in the configuration $\uparrow \uparrow $,  while 
in the arrangement $\downarrow \uparrow$, the roton can be observed only for large $\lambda$. 
The roton instability can be identified by $d\varepsilon_k/dk|_{k=k_r}=0$.
The roton minimum touches zero at $k_r=0$ leading to roton instability. 
Another feature of the spectrum (\ref {spectrum}) is that it is independent of the random potential.

At zero temperature, the noncondensed density is defined as  $\tilde n= \tilde n_0+n_R $, where
\begin{align} \label{NCD1} 
\tilde n_0=\frac{1}{2}\int \frac{d \mathbf k} {(2\pi)^2} \left[\frac{E_k+ gn (1-Ck \mp C k  e^{-k\lambda})} {\varepsilon_k} -1\right], 
\end{align}
accounts for the quantum fluctuations contribution to the noncondensed density. And
\begin{equation}\label{NCD2}
n_R =n\int \frac{d \mathbf k} {(2\pi)^2} R_k \frac{ E_k^2}{\varepsilon_k^4},
\end{equation}
represents the disorder fluctuations to the noncondensed density {\it glassy fraction} analog to the Edwards-Anderson order
parameter of a spin glass  \cite{Edw}. 
It arises from the accumulation of density near the potential minima and density depletion around the maxima.


To understand the interplay of interlayer DDI and disorder effects, let us from now on suppose a correlated Gaussian disorder potential which 
is characterized by two parameters namely :  the disorder strength $R_0$ which has dimension (energy) $^2$ $\times$  (length)$^2$ and 
the correlation length $ \sigma$, it can be written as $R(k)=R_0\exp[-\sigma^2k^2/2]$.

The calculation of integrals (\ref{NCD1}) and (\ref{NCD2})  over infinite momentum space is logarithmically divergent and requires a special care. 
One possibility to solve them is to work with an arbitrary $\Lambda$-cutoff \cite{Jach}. 
However, it turns out that the resulting corrections to the noncondensed density are cutoff-dependent due to the special character of the DDI. 
Another way to treat the above integrals, is the use of a high-momentum cutoff which is 
physically acceptable to obtain qualitatively correct results in the ultracold regime $k \ll 1/r_*$ \cite{Boudj2, Boudj9}.
To be quantitative,  we solve  integrals (\ref{NCD1}) and (\ref{NCD2})  numerically using the standard Monte Carlo method \cite{Boudj2} in the limit $k \ll 1/r_*$. 

\begin{figure}
\includegraphics[scale=0.45]{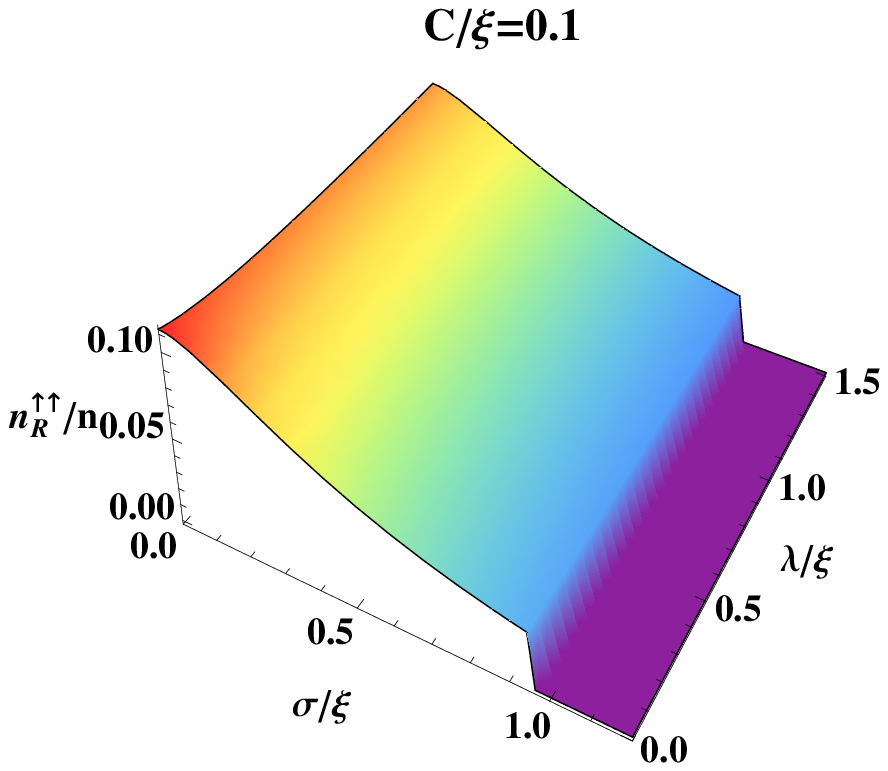}
\includegraphics[scale=0.45]{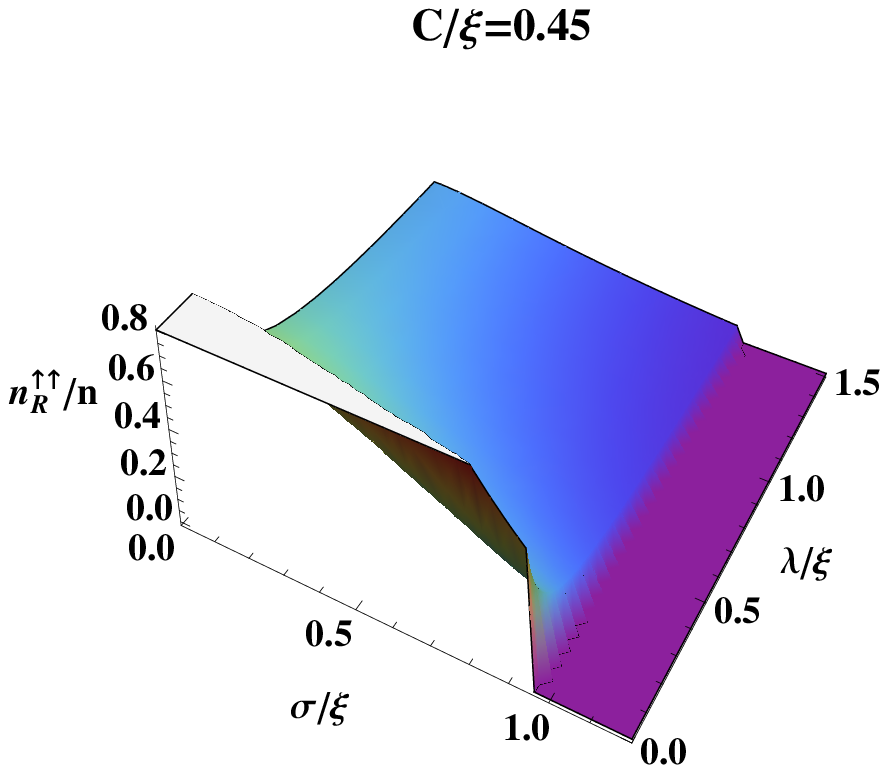}
\includegraphics[scale=0.45]{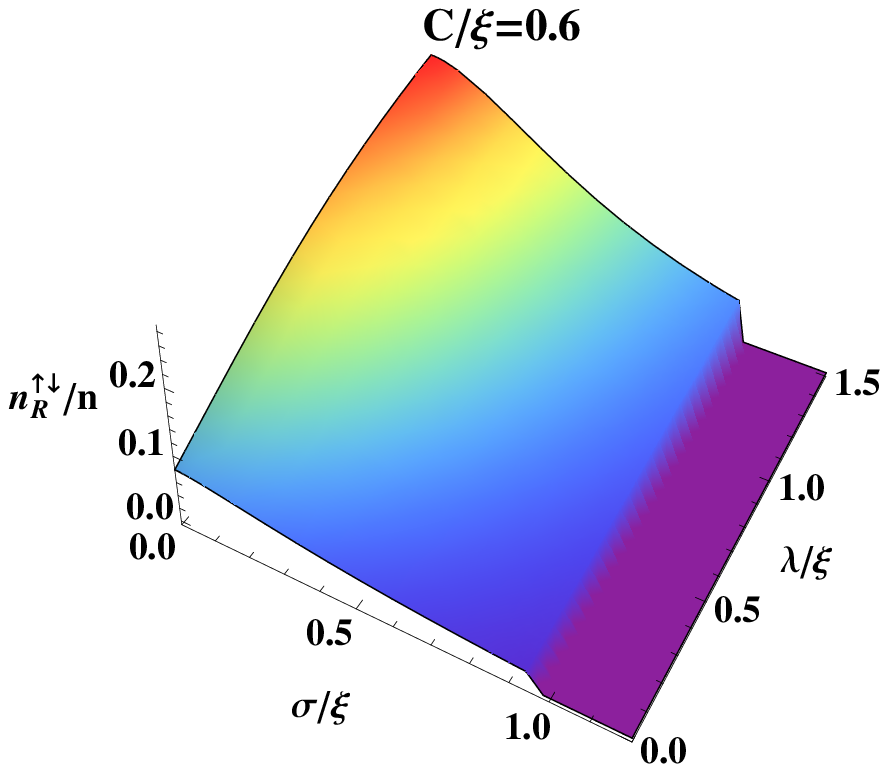}
\includegraphics[scale=0.45]{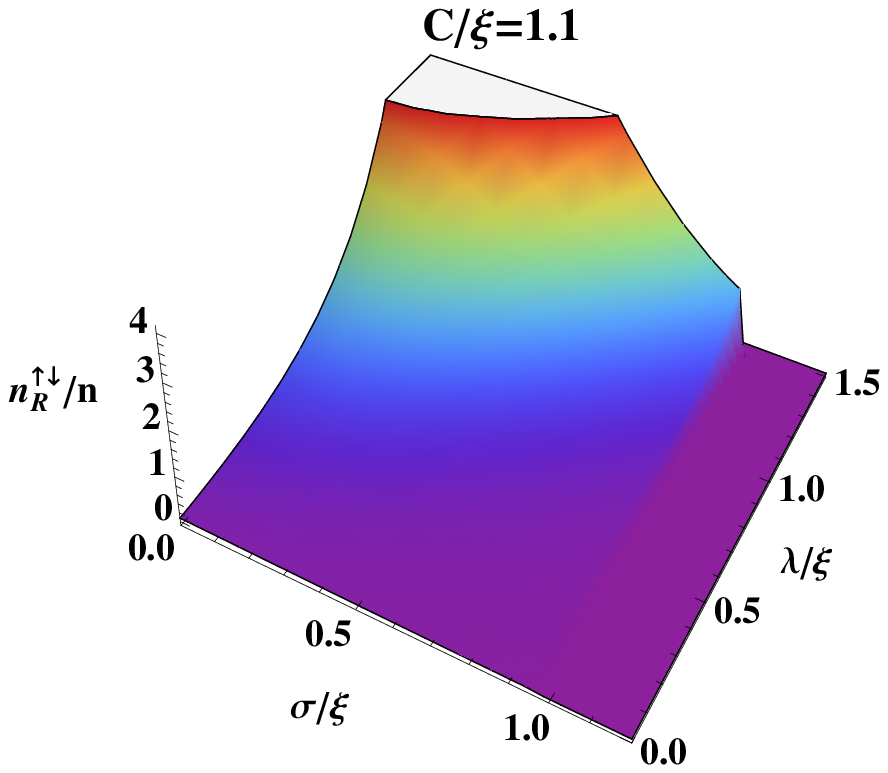}
\includegraphics[scale=0.45]{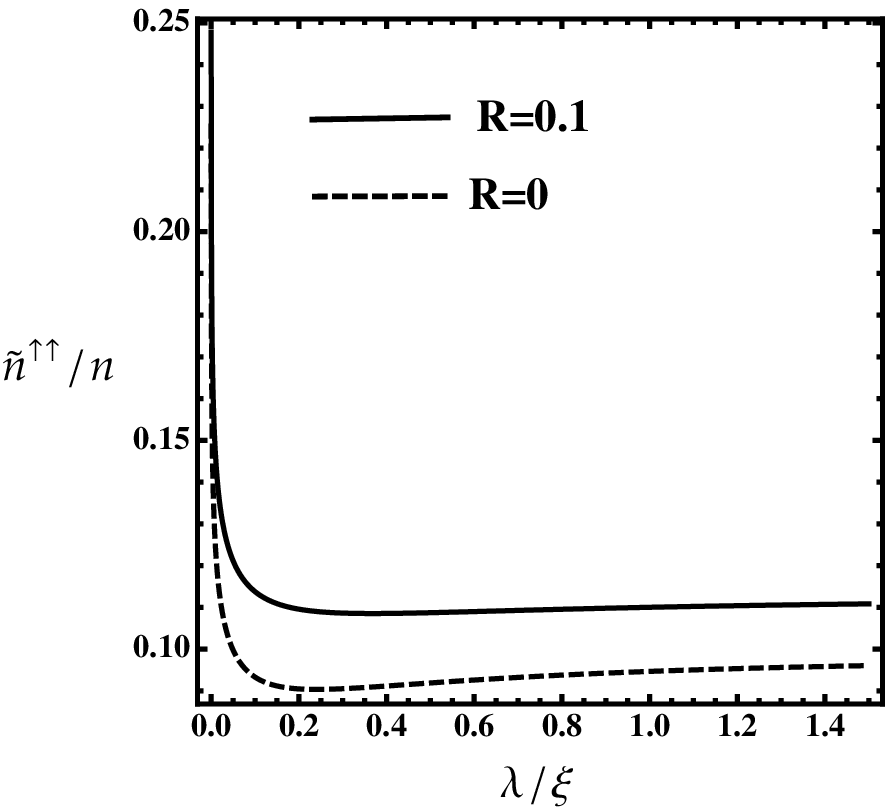}
\includegraphics[scale=0.45]{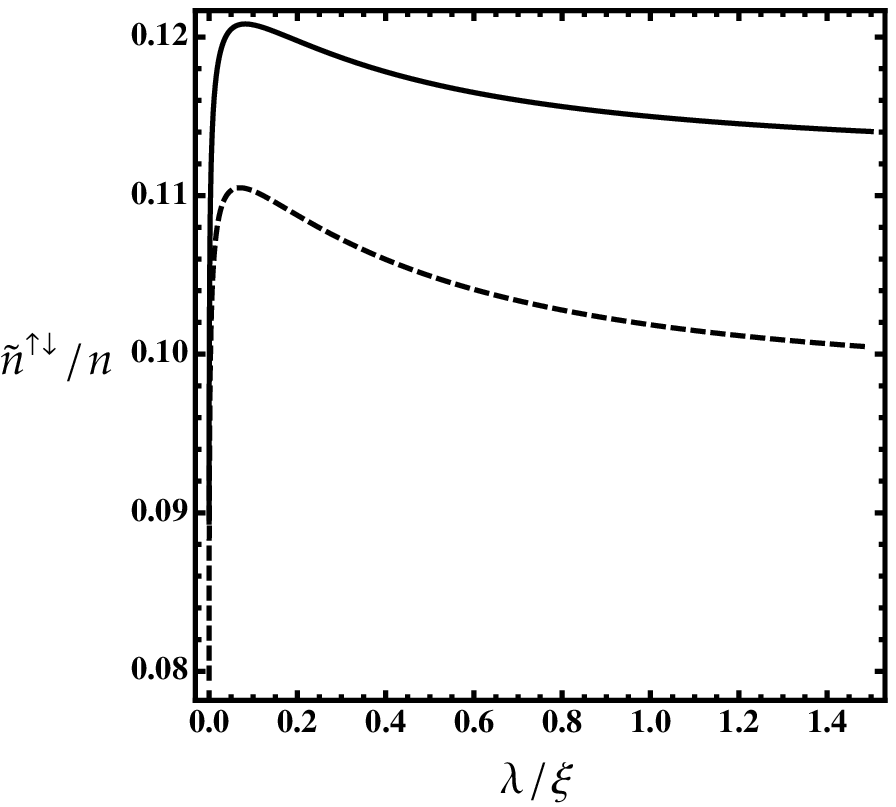}
\caption {(Color online) Glassy fractions $n_{R}^{\uparrow  \uparrow}$ (top panel)  and $n_{R}^{\downarrow  \uparrow}$ (middle panel)  from Eq.(\ref {NCD2})
as a function of $\lambda/\xi$ and $\sigma/\xi$. Parameters are: $mg/4\pi\hbar^2=0.01$ and  R=0.1.
Total condensate depletion as a function of $\lambda/\xi$ for $C/\xi=0.3$ and $\sigma/\xi=0.2$ (bottom panel). 
Here $R=R_0/ng^2$.}
\label{depl}
\end{figure}
Figure.\ref{depl} shows that in the setup $\uparrow \uparrow$, the glassy fraction $n_R$ is decreasing with $\lambda$ indicating that
the interlayer effects lead to tune the disorder fluctuations ensuring the existence of the condensate even for relatively large disorder strength. 
Conversely, in the arrangement $\downarrow \uparrow$,  when the two layers are well separated  ($\lambda/\xi \gtrsim 0.7$), 
$n_R$ substantially increases results in the disappearance of the condensate.
The disorder fraction becomes important when the roton minimum is close to zero (diverges at $k_r=0$)
yielding the transition to a novel quantum phase  \cite {Boudj9} (see right panels).
For $ \sigma >\xi $, the disorder effects is not important in both configurations regardless the polarization directions. 

We observe also that in the absence of the random external potential i.e. $R=0$, the total noncondensed density, $\tilde n^{\uparrow \uparrow}$, lowers
for $\lambda/\xi \lesssim 0.2$ and then grows logarithmically for $\lambda/\xi > 0.2$, where the condensate becomes completely depleted due to the DDI (see bottom panel left).  
However, the situation is inverted in the configuration $\downarrow \uparrow$ (see bottom panel rigth). 
The presence of the disorder potential augments the condensate depletion notably for large $\lambda$ as is seen in the same figure (bottom panel).




\section{Coherence} \label{Coh}

\begin{figure}
\centering{
\includegraphics[scale=0.45]{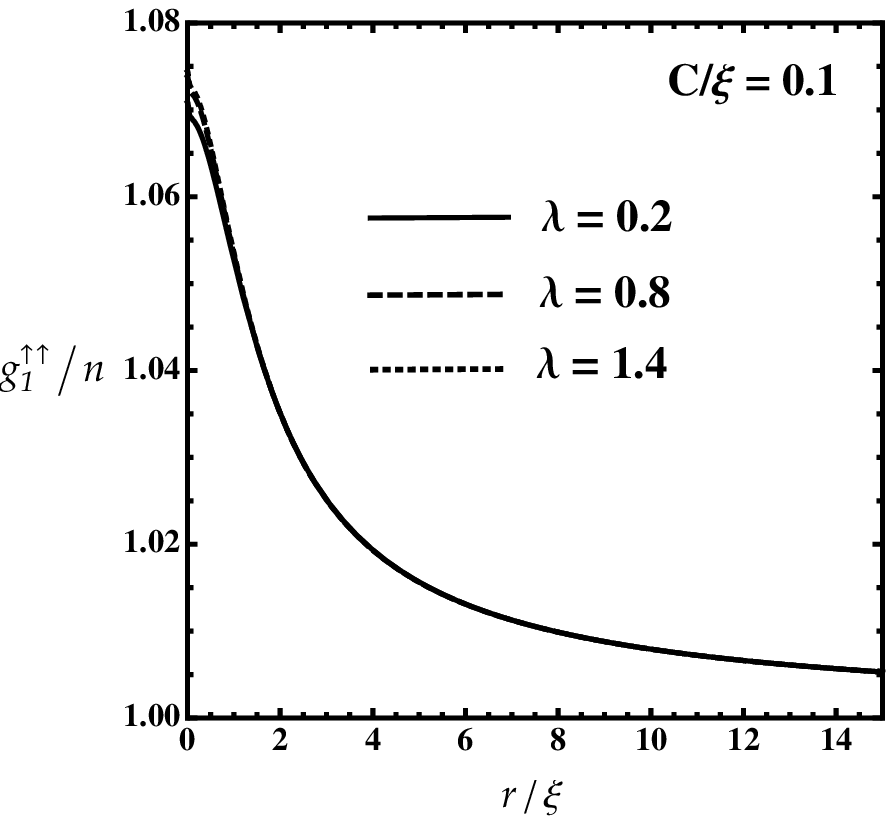}
\includegraphics[scale=0.45]{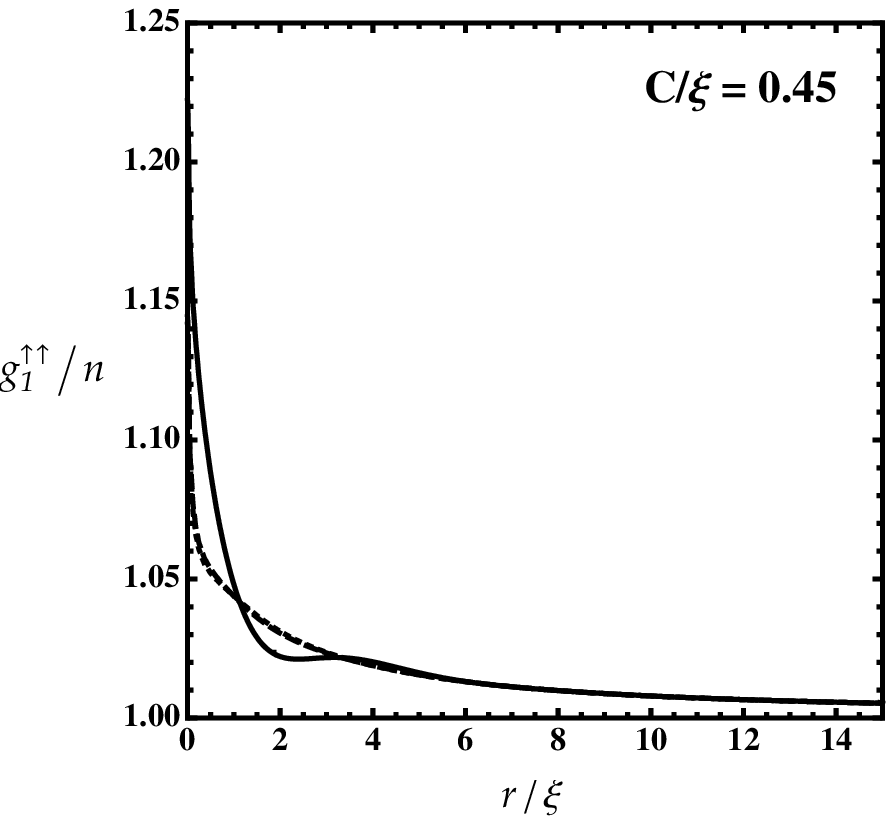}
\includegraphics[scale=0.45]{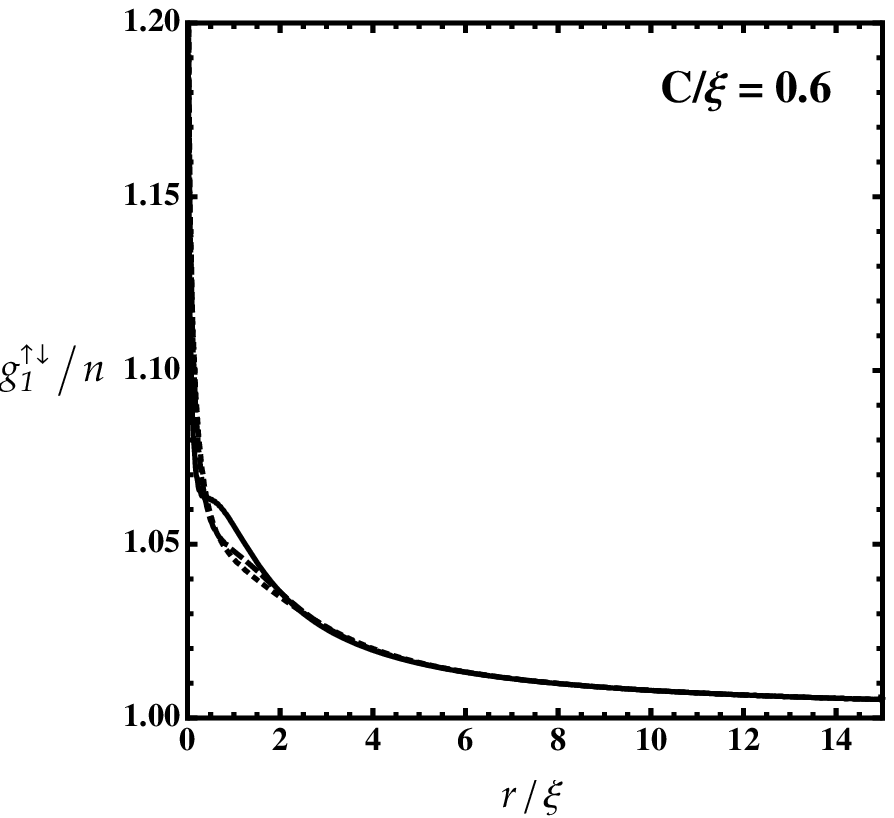}
\includegraphics[scale=0.45]{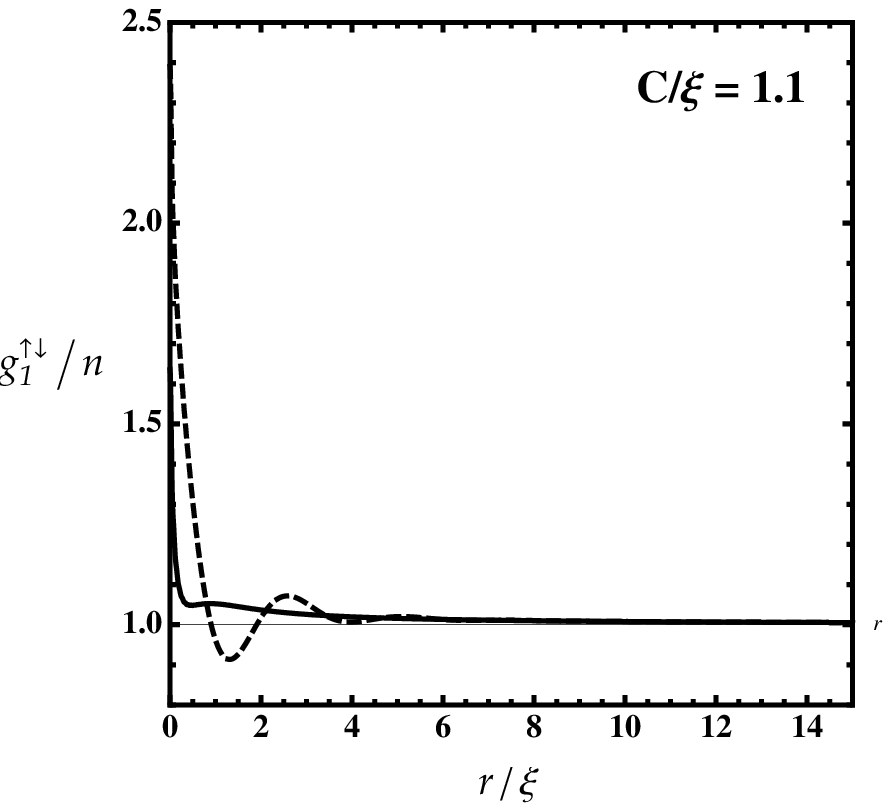}}
\caption {One-body density correlation function for several values of interlayer separation $\lambda$.
Parameters are : $mg/4\pi\hbar^2=0.01$ and $\sigma/\xi=0.2$. }
\label{gd}
\end{figure}

At zero temperature, the one-body density matrix is defined as \cite{Boudj4}
$g_1({\bf r})= n_c+ \int \tilde n \, e^{i \bf k.r} d \mathbf k/(2\pi)^2$, where $n_c$ is the condensed density.
The numerical simulation of this integral reveals that when $C/\xi$ is small, the first order correlation functions
$g_1^{\uparrow \uparrow}({\bf r})$ and $g_1^{\downarrow  \uparrow} ({\bf r})$ decay at large distance and go to their constant value $n$
(see Fig.\ref {gd} left panels). This is a genuine signature of the existence of a true BEC at zero temperature in quasi-2D geometry.
In such a case the interlayer distance and the polarization direction play a minor role; they only slightly shift $g_1^{\uparrow \uparrow}({\bf r})$ 
and $g_1^{\downarrow  \uparrow} ({\bf r})$ near the center. 
For large $C/\xi$ (i.e. when the roton minimum close to zero) and depending on the interlayer space, $g_1^{\uparrow \uparrow}({\bf r})$ and $g_1^{\downarrow  \uparrow} ({\bf r})$ 
display oscillations at small distances (see right panels).
This signals the destruction of the off-diagonal long-range order (i.e., BEC).   
One can conclcude that below a certain critical intralayer coupling $\lambda_c$ which relies on the polarization direction, 
the BEC remains stable. For  $\lambda>\lambda_c$, the system undergoes instability opening the door to a new phase transition.

\section{Superfluidity} \label{superf}

Quasi-2D superfluidity can be well understood in the framework of the Berezinskii-Kosterlitz-Thouless theory \cite{Pop, KT, GPS, Boudj10}.
The relation between the disorder potential, DDI and the superfluidity in quasi-2D geometry has been explained in details in our recent papers \cite{Boudj2, Boudj3, Boudj9}.
The superfluid fraction $n_s/n$ is defined as $n_s/n=1-n_n/n-n_R^{th}/n$ \cite{Boudj2}, where 
\begin{equation}   \label{supp}
 \frac{n_n}{n}=\frac{2}{dTn} \int \frac{d^dk}{(2\pi)^d} \frac{E_k}{4 \text {sinh}^2 (\varepsilon_k/2T)},
\end{equation} 
is the normal fraction of the superfluid.
And 
\begin{equation}   \label{supp1}
 \frac{n_{R\, th}}{n}=\frac{2}{dTn} \int \frac{d^dk}{(2\pi)^d} \frac{n R_k E_k^2}{\varepsilon_k^3}  \text {coth} \left(\varepsilon_k/2T \right),
\end{equation}
represents the disorder thermal contribution to the superfluid fraction.
At temperatures $T \rightarrow 0$, it reduces to $n_{R\,th}/n= 4 n_R/dn$. 
In quasi-2D one has $n_{R\, th}= 2 n_R$ which leads to considerably lower the superfluid fraction.
Another important remark is that the superfluid fraction is no longer a tensorial quantity 
in opposite to the 3D dirty dipolar Bose gas case \cite{Nik, Ghab, Boudj,Boudj1,Boudj2, Boudj3,Boudj4,Boudj5} 
since the dipoles are assumed to be perpendicular to the plane. 
However, in the case of dipolar BECs with tilted dipoles, the superfluid becomes anisotropic.

\begin{figure}
\centering{
\includegraphics[scale=0.45] {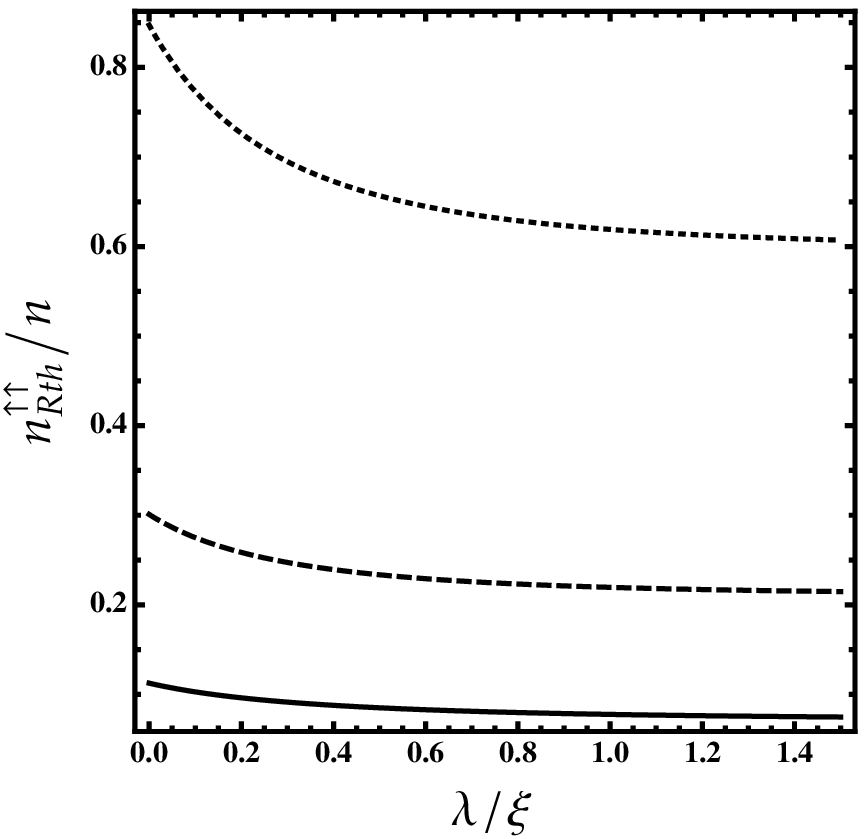}
\includegraphics[scale=0.45]{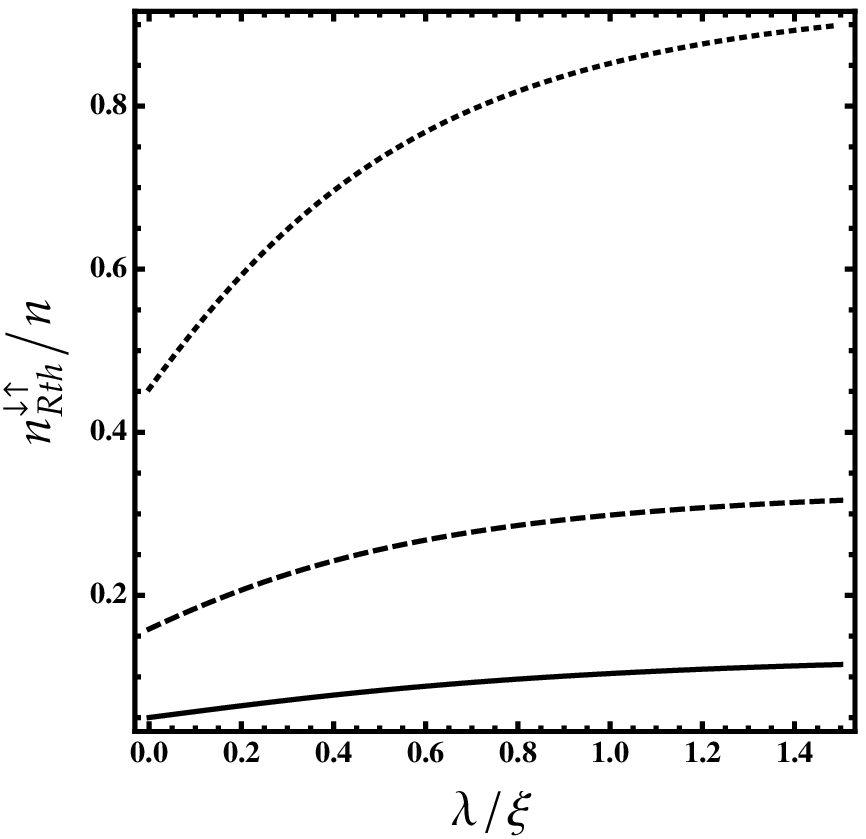}}
\caption {Disorder thermal corrections to the superfluid fraction from Eq.(\ref {supp1}), as a function of $\lambda/\xi$.
Parameters are: $mg/4\pi\hbar^2=0.01$, $\sigma/\xi=0.2$, $C/\xi=0.1$ (left) and $C/\xi=0.6$ (rigth). 
Solid line: $T/ng=0.1$. Dashed line: $T/ng=1$. Dotted line: $T/ng=3$. }
\label{supg}
\end{figure}

Figure.\ref{supg} shows that $n_{R\, th}$ is increasing  with temperature in both configurations.
We see also that $n_{R\, th}^{\uparrow  \uparrow}$ lowers with $\lambda$ at any temperatures. 
For instance, at temperatures $T/ng<0.1$, $n_{R\, th}^{\uparrow  \uparrow} \approx n$ at $\lambda \geq \xi$ which means that
the whole system becomes practically superfluid. 
Whereas, in the configuration ${\downarrow  \uparrow}$, $n_{R\, th}^{\downarrow  \uparrow}$ augments with both temperature and interlayer spacing.
For example, at $T/ng>3$ and for $\lambda > 1.2 \,\xi$, the superfluid fraction vanishes. 
This implies that the condensed particles are localized prohibiting the  superfluid flow results in the formation of the so-called Bose glass phase.

\section{Conclusions} \label{conc}

In this paper  we studied the implications of varying polarization orientation and interlayer DDI
on the propertites of quasi-2D bilayered dipolar Bose gases in a random environment at zero temperature.  
We calculated analytically and numerically  the dispersion relation, the condensate depletion, the first-order correlation function and the superfluidity
using the Bogoliubov theory.
Our analysis revealed that the competition between the disorder potential and the interlayer DDI may significantly enhance the rotonization and the glassy fraction inside the condensate.
In the parallel configuration, interlayer coupling may lead to delocalize atoms rising both the condensed and the superfluid fractions.
However, the situation is completely different in the antiparallel arrangement where the condensate and the superfluid fraction are decreased.
This important result has never been addressed before in the literature.
We showed in addition that in the roton regime, the long-range order is distroyed and hence, the condensate and the superfluidity disappear in both configurations. 
Whereas, for small values of DDI, the coherence of the BEC remains insensitive to interlayer distance.
It was found also that a true BEC exists up to certain critical temperature which depends on the interlayer distance and the polarization direction.
We believe that our results provide new insights to understand these exotic systems, and opening new prospects for realizing dirty dipolar Bose gases.





\begin{thebibliography}{28}

\bibitem{Baranov} See for review: M. A. Baranov, Physics Reports {\bf 464}, 71 (2008).
\bibitem{Pfau} See for review: T. Lahaye et al., Rep. Prog. Phys. {\bf 72}, 126401 (2009).
\bibitem{Carr} See for review: L.D. Carr, D. DeMille, R.V. Krems, and J. Ye, New Journal of Physics {\bf 11}, 055049 (2009).
\bibitem{Pupillo2012} See for review: M.A. Baranov, M. Delmonte, G. Pupillo, and P. Zoller, Chemical Reviews, {\bf 112}, 5012 (2012).

\bibitem{Kobl}  P. K\"oberle, H. Cartarius, T. Fabčič, J. Main, G. Wunner, New J. Phys. {\bf 11}, 023017 (2009).
\bibitem{Mit}  K. A. Mitchell and B. Ilan, Phys. Rev. A80, 043406 (2009).
\bibitem{Rau} S. Rau, J. Main, P. K ̈oberle, and G. Wunner, Phys. Rev. A, {\bf 81}, 031605(R) (2010).
\bibitem{Gut}  R. Gut\"ohrlein, J. Main, H. Cartarius, G. Wunner, J. Phys. A: Math. Theor. {\bf 46},  305001 (2013).

\bibitem{Deng} W. Deng, J. Xu, and H.Zhao, IEEE ACCESS, vol.7, pp.20281-20292, (2019).  
\bibitem{Zhao} H. Zhao,  J. Zheng, J. Xu, and Wu Deng. IEEE ACCESS,  DOI: 10.1109/ACCESS.2019.2929094 (2019). 
\bibitem{Deng2}  W. Deng, H. Zhao, L. Zou, G. Li, X. Yang and D. Wu. Soft Computing, {\bf 21} 4387 (2017). 
\bibitem{Zhao1} H. Zhao, R. Yao, L. Xu, Y. Yuan, G. Li  and W. Deng, Entropy.  {\bf 20}, 682 (2018). 


\bibitem{Clm}  D. Cl\'ement, A. F. Varon, M. Hugbart, J. A. Retter, P. Bouyer, L. Sanchez-Palencia, D. M. Gangardt, G. V. Shlyapnikov, and A. Aspect, Phys. Rev. Lett. {\bf 95}, 170409 (2005).
\bibitem{Schut}  T. Schulte, S. Drenkelforth, J. Kruse, W. Ertmer, J. Arlt, K. Sacha, J. Zakrzewski, and M. Lewenstein,  Phys. Rev. Lett. {\bf 95}, 170411 (2005).
\bibitem{Lye}  J. E. Lye, L. Fallani,M.Modugno, D. S.Wiersma, C. Fort, and M. Inguscio, Phys. Rev. Lett. {\bf 95}, 070401 (2005).
\bibitem{Clm1}  D. Cl\'ement et {\textit al}., New J. Phys. {\bf 8}, 165 (2006).
\bibitem{Bily} J. Billy, V. Josse, Z. Zuo, A. Bernard, B. Hambrecht, P. Lugan, D. Cl\'ement, L. Sanchez-Palencia, P. Bouyer, and A. Aspect, Nature {\bf 453}, 891 (2008).
\bibitem{Roat}  G. Roati, C. D.Errico, L. Fallani, M. Fattori, C. Fort, M. Zaccanti, G. Modugno, M. Modugno, and M. Inguscio,  Nature {\bf 453}, 895 (2008).
\bibitem{Chen} Y. P. Chen, J. Hitchcock, D. Dries, M. Junker, C. Welford, and R. G. Hulet, Phys. Rev. A {\bf 77}, 033632 (2008).
\bibitem{Wht}  M. White, M. Pasienski, D. McKay, S. Q. Zhou, D. Ceperley, and B. DeMarco,  Phys. Rev. Lett. {\bf 102}, 055301 (2009).
\bibitem{HM}  K. Huang and H.-F. Meng, Phys. Rev. Lett. {\bf 69}, 644 (1992).
\bibitem{Gior}  S. Giorgini, L. Pitaevskii, and S. Stringari, Phys. Rev. B {\bf 49}, 12 938 (1994).
\bibitem{Mish} M. Kobayashi and M. Tsubota, Phys. Rev. B {\bf 66}, 174516 (2002).
\bibitem{Lugan2} P. Lugan, D. Cl\'ement, P. Bouyer, A. Aspect, M. Lewenstein, and L. Sanchez-Palencia,  Phys. Rev. Lett. {\bf 98}, 170403 (2007).
\bibitem{Falco} G. M. Falco, A. Pelster, and R.Graham, Phys.Rev.A {\bf 75}, 063619 (2007).
\bibitem{Yuk} V. I. Yukalov and R. Graham, Phys.Rev.A {\bf 75}, 023619 (2007).
\bibitem{Lopa} A.V. Lopatin and V. M. Vinokur, Phys. Rev. Lett. {\bf 88}, 235503 (2002).
\bibitem{Zob} O. Zobay, Phys. Rev. A {\bf 73}, 023616 (2006).
 \bibitem{Mor} S. Morrison, A. Kantian, A. J. Daley, H. G. Katzgraber, M. Lewenstein, H. P. Buechler, and P. Zoller, New J. Phys. {\bf 10}, 073032 (2008).
 \bibitem{Bhong} S. G. Bhongale, P. Kakashvili, C. J. Bolech, and H. Pu, Phys. Rev. A {\bf 82}, 053632 (2010).
\bibitem{LSP} L. Sanchez-Palencia, Phys. Rev. A {\bf 74}, 053625 (2006).
\bibitem{Lugan} P. Lugan and L. Sanchez-Palencia, Phys. Rev. A {\bf 84}, 013612 (2011).
\bibitem{Lugan1} P. Lugan, D. Cl\'ement, P. Bouyer, A. Aspect, and L. Sanchez-Palencia, Phys. Rev. Lett. {\bf 99}, 180402 (2007).
\bibitem{Gaul} C. Gaul and C. A. M\"uller, Phys. Rev. A {\bf 83}, 063629 (2011).
\bibitem{Gaul1} C. A. M\"uller and C. Gaul, New. J. Phys {\bf 14},  075025 (2012).
\bibitem{Lell} S. Lellouch, L-K Lim, and L. Sanchez-Palencia, Phys. Rev. A {\bf 92}, 043611 (2015).

\bibitem{Krum} C. Krumnow and A. Pelster, Phys. Rev. A {\bf 84}, 021608(R) (2011).
\bibitem{Nik} B. Nikolic,  A. Balaz, and A. Pelster, Phys. Rev. A {\bf 88}, 013624 (2013).
\bibitem{Ghab}  M. Ghabour and A. Pelster, Phys. Rev. A {\bf 90}, 063636 (2014).
\bibitem{Boudj} A. Boudjem\^{a}a, Phys. Rev. A {\bf 91}, 053619 (2015).
\bibitem{Boudj1} A. Boudjem\^{a}a,  Low Temp. Phys. {\bf 180}, 377 (2015).
\bibitem{Boudj2} A. Boudjem\^{a}a, Phys. Lett. A {\bf 379}, 2484 (2015).
\bibitem{Boudj3} A. Boudjem\^{a}a,  J. Phys. B: At. Mol. Opt. Phys. {\bf 49}, 105301 (2016).
\bibitem{Boudj4} K. Redaouia and  A. Boudjem\^{a}a, Eur. Phys. J. D {\bf 73},115 (2019).
\bibitem{Boudj5} A. Boudjem\^{a}a, Eur. Phys. J. B {\bf 92},145  (2019).


\bibitem {Wang} D.-W. Wang, M. D. Lukin, and E. Demler, Phys. Rev. Lett. {\bf 97}, 180413 (2006).
\bibitem {Wang1} D.-W. Wang and E. Demler, arXiv:0812.1838.
\bibitem {Piko} A. Pikovski, M. Klawunn, G. V. Shlyapnikov, and L. Santos, Phys. Rev. Lett. {\bf 105}, 215302 (2010).
\bibitem {Misha} M. A. Baranov, A. Micheli, S. Ronen, and P. Zoller, Phys. Rev. A {\bf 83}, 043602 (2011).
\bibitem {Shi} S.-M. Shih and D.-W. Wang, Phys. Rev. A {\bf 79}, 065603 (2009).
\bibitem {Pot} A. C. Potter, E. Berg,D.-W.Wang, B. I. Halperin, and E. Demler, Phys. Rev. Lett. {\bf 105}, 220406 (2010).
\bibitem {Vol} A. G. Volosniev, D. V. Fedorov, A. S. Jensen, and N. T. Zinner, Phys. Rev. Lett. {\bf 106}, 250401 (2011).
\bibitem {Dalm}  M. Dalmonte, P. Zoller, G. Pupillo, Phys. Rev. Lett. {\bf 107}, 163202 (2011).
\bibitem{Santos1} R. Nath, P. Pedri, and L. Santos, Phys. Rev. A {\bf 76}, 013606 (2007).
\bibitem{Santos2} R. Nath, P. Pedri, and L. Santos, Phys. Rev. A {\bf 86}, 013610 (2012).
\bibitem{Ros} M. Rosenkranz, Y. Cai, and W. Bao, Phys. Rev. A {\bf 88}, 013616 (2013).
\bibitem{Klm} M. Klawunn and A. Recati, Phys. Rev. A {\bf 88}, 013633 (2013).
\bibitem{Fedo} A.K. Fedorov, S.I. Matveenko, V.I. Yudson, G.V. Shlyapnikov,  Sci. Rep. {\bf 6},  27448 (2016).
\bibitem{Boudj6} A.Boudjem\^aa,  Phys. Lett. A {\bf 381}, 1745 (2017).
\bibitem{Boudj7} A.Boudjem\^aa,  J. Low. Temp. Phys {\bf 189}, 76 (2017).
\bibitem{Boudj8} K. Mohammed Elhadj, A. Boudjem\^{a}a, and U. Al-Khawaja, Phys. Scr. {\bf 94}, 085402 (2019).

\bibitem{Boudj9}  A. Boudjem\^{a}a, and G.V. Shlyapnikov, Phys. Rev. A {\bf 87}, 025601 (2013).
\bibitem{Edw} S. F. Edwards and P.W. Anderson,   J. Phys. F {\bf 5}, 965 (1975).
\bibitem{Jach}  K. Jachymski and R. Ołdziejewski, Phys. Rev. A {\bf 98}, 043601 (2018).
\bibitem{Pop} V.N. Popov, {\it Functional Integrals in Quantum Field Theory and Statistical Physics} (D. Reidel Pub., Dordrecht, 1983).
\bibitem{Boudj10} A. Boudjem\^{a}a, Phys. Rev. A {\bf 86}, 043608 (2012).
\bibitem{KT} J.M. Kosterlitz and D.J. Thouless, J.Phys. C {\bf 6}, 1181 (1973); J.M. Kosterlitz, J. Phys. C {\bf 7}, 1046 (1974).
\bibitem{GPS} See for review: D.S. Petrov, D.M. Gangardt, and G.V. Shlyapnikov, J. Phys. IV (France) {\bf 116}, 5 (2004).




\end{thebibliography}
\end{document}